\documentclass[letterpaper, 10 pt, conference]{ieeeconf}

\IEEEoverridecommandlockouts                              % This command is only
\usepackage{amsmath}    % need for sub equations
\usepackage{amsfonts}
\usepackage{graphicx}   % need for figures
\usepackage{subcaption}
\usepackage{epsfig} 
\usepackage{cancel}
\usepackage{amssymb}
\usepackage{color}
\usepackage{cite}
\usepackage{bm}
\usepackage[ruled,vlined,titlenotnumbered,linesnumbered]{algorithm2e} 
\usepackage{todonotes} 
\usepackage{float}
\usepackage{dsfont}			%for the symbol 'Real Numbers'
\usepackage{mathtools}
\usepackage{wrapfig}

\newcommand{\R}{\mathbb{R}} % Real number
\newcommand{\state}{z} % state
\newcommand{\ctrl}{u} % control
\newcommand{\dyn}{f} % dynamics
\newcommand{\thetaReg}{\mathcal{M}} %Valid region for theta
\newcommand{\pol}{\pi} %Policy for linear system
\newcommand{\stateseq}{\bold{\state}} %State sequence
\newcommand{\ctrlseq}{\bold{\ctrl}} %State sequence
\newcommand{\param}{\theta}
\newcommand{\intsymbol}{'}
\newcommand{\paramA}[1]{A_{\param^{#1}}}
\newcommand{\paramB}[1]{B_{\param^{#1}}}
\newcommand{\metName}{\normalfont{a}DOBO}
\newcommand{\cost}{J}
\newcommand{\pos}{p} % position
\newcommand{\compMet}{\eta}
\newcommand{\squeezeup}[1]{\vspace{#1}}

\newcommand{\tab}[1]{Table~\ref{#1}}
\newcommand{\alg}[1]{Algorithm~\ref{#1}}

\newcommand{\citep}[1]{\cite{#1}} % If not using bibnat

\newcommand{\capitalize}[1]{\expandafter\MakeUppercase\expandafter{#1}}

\renewcommand{\vec}[1]{\boldsymbol{#1}}				% Vector
\newcommand{\mat}[1]{\boldsymbol{\capitalize{#1}}}		% Matrix
\DeclareMathOperator{\diag}{\mathrm{diag}} 			% Diagonal matrix

\newcommand{\dataset}[0]{\mathcal{D}}

\newcommand{\parameter}[0]{\theta} 				% Parameter
\newcommand{\parameters}[0]{{\parameter}} 			% Parameters
\newcommand{\dimparameters}{d}					% Dimensionality parameters
\newcommand{\objfuncNo}[0]{J}					% Objective function
\newcommand{\objfunc}[1]{\objfuncNo\left(#1\right)}		% Objective function (as a function)

\newcommand{\acqfuncNo}[0]{\alpha}				% Acquisition function
\newcommand{\acqfunc}[1]{\acqfuncNo \left( #1 \right)}		% Acquisition function (as a function)
\title{\LARGE \bf
Goal-Driven Dynamics Learning via Bayesian Optimization}
\author{Somil Bansal, Roberto Calandra, Ted Xiao, Sergey Levine, and Claire J. Tomlin
\thanks{All authors are with the Department of Electrical Engineering and Computer Sciences, University of California, Berkeley. \{somil, roberto.calandra, t.xiao, svlevine, tomlin\}@eecs.berkeley.edu}
\thanks{${}^*$This research is supported by NSF under the CPS Frontiers VehiCal project (1545126), by the UC-Philippine-California Advanced Research Institute under project IIID-2016-005, and by the ONR MURI Embedded Humans (N00014-16-1-2206).}
}

\begin{document}
\maketitle
\thispagestyle{empty}
\pagestyle{empty}

%%%
\begin{abstract}
Real-world robots are becoming increasingly complex and commonly act in poorly understood environments where it is extremely challenging to model or learn their true dynamics. 
Therefore, it might be desirable to take a task-specific approach, wherein the focus is on explicitly learning the dynamics model which achieves the best control performance for the task at hand, rather than learning the true dynamics. 
In this work, we use Bayesian optimization in an active learning framework where a locally linear dynamics model is learned with the intent of maximizing the control performance, and used in conjunction with optimal control schemes to efficiently design a controller for a given task. 
This model is updated directly based on the performance observed in experiments on the physical system in an iterative manner until a desired performance is achieved.  
We demonstrate the efficacy of the proposed approach through simulations and real experiments on a quadrotor testbed.   
\end{abstract}
\setlength{\textfloatsep}{10pt}

% !TEX root = SysIdLQR.tex
\section{Introduction}
Given the system dynamics, optimal control schemes such as LQR, MPC, and feedback linearization can efficiently design a controller that maximizes a performance criterion. 
However, depending on the system complexity, it can be quite challenging to model its true dynamics. 
Moreover, for a given task, a globally accurate dynamics model is not always necessary to design a controller.
Often, partial knowledge of the dynamics is sufficient, e.g., for trajectory tracking purposes a local linearization of a non-linear system is often sufficient. 
In this paper we argue that, for complex systems, it might be preferable to adapt the controller design process for the specific task, using a learned system dynamics model sufficient to achieve the desired performance. 

\begin{figure}[t!]
  \centering
  \includegraphics[width=0.45\textwidth]{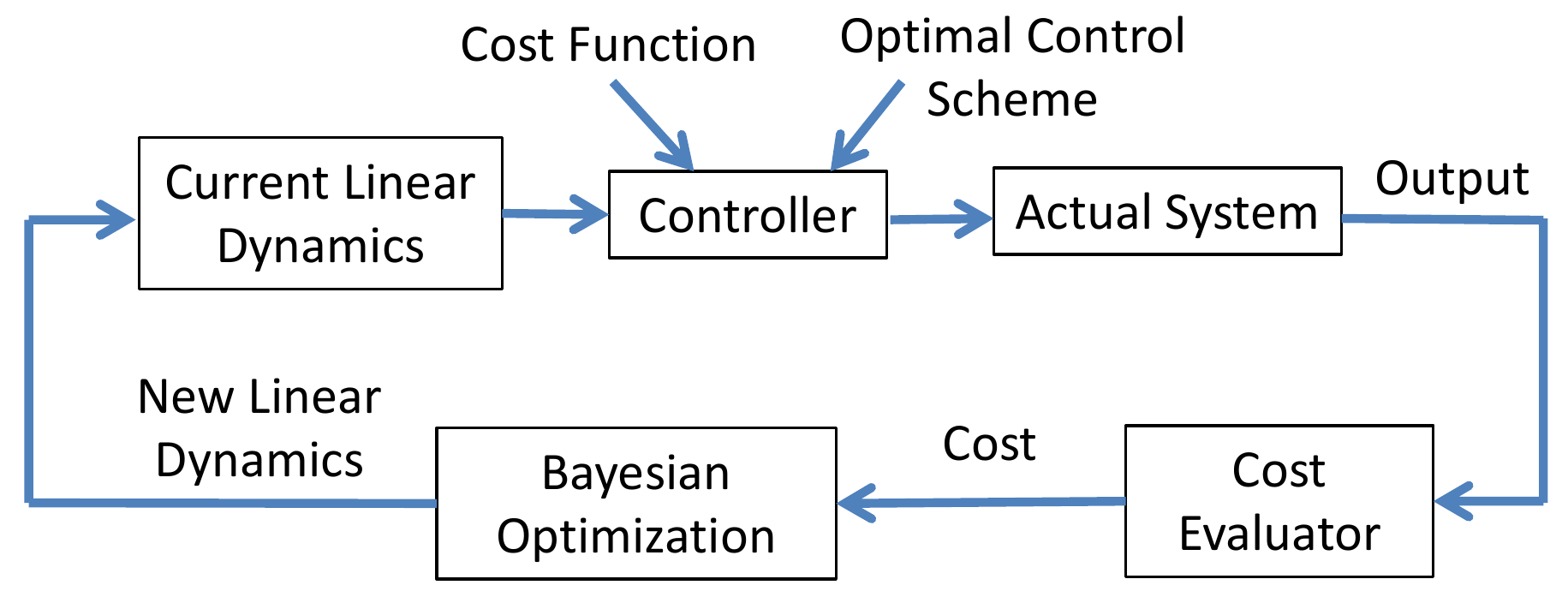}
  \caption{\metName: A Bayesian optimization-based active learning framework for optimizing the dynamics model for a given cost function, directly based on the observed cost values.}
  \label{fig:framework}
  \vspace{-8pt}
\end{figure}

We propose Dynamics Optimization via Bayesian Optimization (\metName), a Bayesian Optimization (BO) based active learning framework to learn the dynamics model that achieves the best performance for a given task based on the performance observed in experiments on the physical system. 
This active learning framework takes into account all past experiments and suggests the next experiment in order to learn the most about the relationship between the performance criterion and the model parameters. 
Particularly important for robotic systems is the use of data-efficient approaches, where only few experiments are needed to obtain improved performance. 
%To make sure our approach is data-efficient, we learn a \textit{locally linear} system model and use that to design the controller, as opposed to learning a global model or learning control policy directly. The underlying hypothesis being that a good local model in conjunction with well analyzed optimal control schemes can be used to design a controller more efficiently, and can capture a richer controller space.Specifically, we use Bayesian optimization for actively learning the local dynamics. 
Hence, we employ BO, an optimization method often used to optimize a performance criterion while keeping the number of evaluations of the physical system small~\cite{Shahriari2016}.
%, e.g., when an evaluation requires an expensive interaction with a system. 
%BO has been used in the past to directly optimize the gains of a controller~\cite{Calandra2015a} or the cost function of a LQR controller~\cite{Marco2016}. 
Specifically, we use BO to optimize the dynamics model with respect to the desired task, where the dynamics model is updated after every experiment so as to maximize the performance on the physical system. 
A flow diagram of our framework is shown in Figure~\ref{fig:framework}. 
The current linear dynamics model, together with the cost function (also referred to as task/performance criterion), are used to design a controller with an appropriate optimal control scheme. 
The cost (or performance) of the controller is evaluated in closed-loop operation with the actual (unknown) physical plant. 
BO uses this performance information to iteratively update the dynamics model to improve the performance. 
This procedure corresponds to optimizing the linear system dynamics with the purpose of maximizing the performance of the final controller. 
Hence, unlike traditional system identification approaches, it does not necessarily correspond to finding the most accurate dynamics model, but rather the model yielding the best controller performance when provided to the optimal control method used. 
%This objective sets apart our approach from traditional system identification approaches, where the focus is on learning an accurate prediction model, as opposed to maximizing the controller performance. 
% In contrast to traditional system identification approaches, where the focus is on learning an accurate prediction model, our approach aim to directly maximize the controller performance.
% Our approach differ from the traditional system identification approaches, where the focus is on learning an accurate prediction model, as it directly maximize the controller performance. 
%An interesting question to study in the context of this method is the degree to which the resulting model actually corresponds to the true dynamics of the system. We study this in Section~\ref{sec:simple1D}.
%
% 

%Traditional system identification approaches are used in control through a two-stage design process clearly divided into: 1) creating a dynamics model by minimizing some prediction error (e.g., using least squares) 2) using this dynamics model to generate an appropriate controller. 
%In this process, modeling the dynamics can be considered an offline process as there is no information flow between the two design stages.
%In online methods instead, the dynamics model is iteratively updated using new data collected by evaluating the controller~\cite{Deisenroth2015}.
%Our approach belongs to the online methods.
%Both for the online and the offline cases, even small inaccuracies in the dynamics model can significantly deteriorate the performance of the overall control system.
%Using machine learning techniques, such as Gaussian processes, do not alleviate this issues~\citep{Nguyen-Tuong2011}.
%\cite{Joseph2013} demonstrated for the offline case that creating a dynamics model by only minimizing the prediction error can introduce sufficient inaccuracies to lead to suboptimal performance.
%Instead, they proposed to optimize the dynamics model directly w.r.t the controller performance. 
Traditional system identification approaches are divided into two stages: 1) creating a dynamics model by minimizing some prediction error (e.g., using least squares) 2) using this dynamics model to generate an appropriate controller. 
In this approach, modeling the dynamics can be considered an offline process as there is no information flow between the two design stages.
In online methods, the dynamics model is instead iteratively updated using the new data collected by evaluating the controller~\cite{Deisenroth2015}. 
Our approach is an online method.
Both for the online and the offline cases, creating a dynamics model based only on minimizing the prediction error can introduce sufficient inaccuracies to lead to suboptimal control performance~\cite{Joseph2013, donti2017task, atkeson1998nonparametric, abbeel2006using, gevers2005identification, hjalmarsson1996model}.
Using machine learning techniques, such as Gaussian processes, does not alleviate this issue~\citep{Nguyen-Tuong2011}. 
Instead, authors in \cite{Joseph2013} proposed to optimize the dynamics model directly with respect to the controller performance, but since the dynamics model is optimized offline, the resultant model is not necessarily optimal for the actual system. 
We instead explicitly find the dynamics model that produces the best control performance for the \emph{actual} system. 

Previous studies addressed the problem of optimizing a controller using BO. 
In \cite{Marco2016, trimpe2014self, roberts2011feedback}, the authors tuned the penalty matrices in an LQR problem for performance optimization. 
%Although interesting results emerge from these studies, it is not clear how these methods perform for non-quadratic cost functions. 
%Moreover, when an accurate system model is not available, tuning penalty matrices may not achieve the desired performance. 
Although interesting results emerge from these studies, tuning penalty matrices may not achieve the desired performance when an accurate system model is not available. 
Our approach overcomes these challenges as it does not rely on an accurate system dynamics model.
In \cite{Calandra2015a}, the authors directly learn the parameters of a linear feedback controller using BO.
%Parameters of a linear feedback controller are directly learned in \cite{Calandra2015a} using BO. 
However, a typical controller might be non-linear and can contain hundreds of parameters; it is not feasible to optimize such high-dimensional controllers using BO~\cite{Shahriari2016}. 
%Our approach overcomes these challenges as it does not rely on an accurate system dynamics model or impose any linear structure on the controller. 
Since aDOBO does not aim at directly learning the controller, it is agnostic to the dimensionality of the controller. 
It can leverage the low-dimensional structure of the dynamics to optimize the high-dimensional controllers.
Moreover, it does not need to impose any structure on the controller and can easily design general non-linear controllers as well (see Sec. \ref{solution}). 
%For example, for non-quadratic convex cost functions, we can use MPC as the optimal control scheme to design non-linear controllers. %and can automatically design a controller even when the system dynamics are unknown and/or the performance criterion is not quadratic. 
%
The problem of updating a system model to improve control performance is also related to adaptive control, where the model parameters are identified from sensor data, and subsequently the updated model is used to design a controller (see \cite{aastrom2013adaptive, grimble1984implicit, clarke1985generalized, murray2002nonlinear, sastry2011adaptive}). 
However, in adaptive control, the model parameters are generally updated to get a good prediction model and not necessarily to maximize the controller performance. 
In contrast, we explicitly take into account the observed performance and search for the model that achieves the highest performance. 

To the best of our knowledge, this is the first method that optimizes a dynamics model to maximize the control performance on the actual system. 
Our approach does not require the prior knowledge of an accurate dynamics model, nor of its parameterized form. 
Instead, the dynamics model is optimized, in an active learning setting, directly with respect to the desired cost function using data-efficient BO.
%The contribution of this paper is to present an automatic approach to controller design. This approach does not require the prior knowledge of an accurate dynamics model, nor of its parameterized form. 
%Instead, the dynamics model is optimized, in an active learning setting, directly w.r.t. the desired cost function using data-efficient Bayesian optimization.
%The idea of using a model that is not the most likely model, but rather the one that achieves the best expected reward has been previously proposed in \cite{Joseph2013}. Although similar in spirit, their approach rely on a gradient-descent optimizer which requires to numerically approximate the gradient by central difference, resulting in an increase in the number of experiments required. Our approach instead, makes use of a global zero-order optimizer (i.e., it does not need gradients) and can therefore be more sample efficient. We further compare some of these approaches with the proposed approach in \sec{sec:comparison}. To summarize, our main contributions in this work are:
%\begin{itemize}
%\item to efficiently and automatically design a controller for general performance criterion, even when the system dynamics are unknown;  
%\item to compare different automatic controller design approaches and highlight their relative advantages and limitations.  
%\end{itemize}
% Introduction (1-1.5p)
%% Motivation
%% Related work
%% Summary of results
%
% !TEX root = SysIdLQR.tex
\section{Problem Formulation} \label{sec:formulation}
Consider an unknown, stable, discrete-time, potentially non-linear, dynamical system
\begin{equation} \label{eqn:sysmodel}
\state_{k+1} = \dyn(\state_k, \ctrl_k),\quad k \in \{0, 1, \ldots, N-1\}\,,
\end{equation}
where $\state_k \in \R^{n_x}$ and $\ctrl_k \in \R^{n_u}$ denote the system state and the control input at time $k$ respectively. Given an initial state $\state_0$, the objective is to design a controller that minimizes the cost function $\cost$ subject to the dynamics in \eqref{eqn:sysmodel}
\begin{equation} \label{eqn:nl_ctrl}
\begin{aligned} 
\cost_0^* = \min_{\ctrlseq_0^{N-1}}\cost_0(\stateseq_0^N, \ctrlseq_0^{N-1}) = & \min_{\ctrlseq_0^{N-1}}\sum_{i=0}^{N-1} l(\state_i, \ctrl_i) + g(\state_N, \ctrl_N)\,,\\
\text{subject to}~~ \state_{k+1} = & \dyn(\state_k, \ctrl_k)\,,
\end{aligned}
\end{equation}
where $\stateseq_i^N := (\state_i, \state_{i+1}, \ldots, \state_N)$. $\ctrlseq_i^{N-1}$ is similarly defined. One of the key challenges in designing such a controller is the modeling of the unknown system dynamics in \eqref{eqn:sysmodel}. 
In this work, we model \eqref{eqn:sysmodel} as a linear time-invariant (LTI) system with system matrices~$(\paramA{},\paramB{})$. 
The system matrices are parameterized by $\param \in \thetaReg \subseteq \R^\dimparameters$, which is to be varied during the learning procedure. 
For a given $\param$ and the current system state $\state_k$, let $\pol_k(\state_k, \param)$ denote the optimal control sequence for the \textit{linear} system $(\paramA{},\paramB{})$ for the horizon $\{k, k+1, \ldots, N\}$
\begin{equation} \label{eqn:lin_ctrl}
\begin{aligned} 
\pol_k(\state_k, \param) :=  \bold{\bar{\ctrl}}_k^{N-1} =  & \arg \min_{\ctrlseq_k^{N-1}} J_k(\stateseq_k^N, \ctrlseq_k^{N-1})\,,\\
\text{subject to}~~ \state_{j+1} = & \paramA{} \state_j + \paramB{} \ctrl_j.
\end{aligned}
\end{equation}
The key difference between \eqref{eqn:nl_ctrl} and \eqref{eqn:lin_ctrl} is that the controller is designed for the parameterized linear system as opposed to the true system. 
As $\param$ is varied, different matrix pairs $(\paramA{},\paramB{})$ are obtained, which result in different controllers~$\pol(\cdot, \param)$. 
Our aim is to find, among all linear models, the linear model $(\paramA{*},\paramB{*})$ whose controller $\pol(\cdot, \param^*)$ minimizes $\cost_0$ (ideally achieves $\cost_0^*$) for the \textit{actual system}, i.e., 
%Mathematically, we want to solve the optimization problem
%
\begin{equation} \label{eqn:optzProb}
\begin{aligned} 
\param^* = & \arg \min_{\param \in \thetaReg} \cost_0(\stateseq_0^N, \ctrlseq_0^{N-1})\,, \\
\text{subject to}~~ & \state_{k+1} = \dyn(\state_k, \ctrl_k)\,, \quad \ctrl_k = \pol_k^1(\state_k, \param),
\end{aligned}
\end{equation} 
where $\pol_k^1(\state_k, \param)$ denotes the $1$st control in the sequence~$\pol_k(\state_k, \param)$. 
To make the dependence on $\param{}$ explicit, we refer to $\cost_0$ in \eqref{eqn:optzProb} as $\cost(\param{})$ here on. 
%\eqref{eqn:optzProb} aims to find the linear model that gives the best performance on the actual system when its optimal controller is applied in a \textit{closed-loop} fashion on the actual physical plant. 
%However, $(\paramA{*},\paramB{*})$ may not correspond to an actual linearization of the system, but simply to the linear model that minimizes the cost function best. 
Note that $(\paramA{*},\paramB{*})$ in \eqref{eqn:optzProb} may not correspond to an actual linearization of the system, but simply to the linear model that gives the best performance on the actual system when its optimal controller is applied in a \textit{closed-loop} fashion on the actual physical plant.   

We choose LTI modeling to reduce the number of parameters used to represent the system, and make the dynamics learning process data efficient. 
%, which is a critical consideration in any active learning approach, as the interaction with the actual system is often expensive. 
Linear modeling also allows to efficiently design the controller in \eqref{eqn:lin_ctrl} for general cost functions (e.g., using MPC for any convex cost~$\cost$). 
In general, the effectiveness of linear modeling depends on both the system and the control objective. 
If $\dyn$ is linear, a linear model is trivially sufficient for any control objective. 
If $\dyn$ is non-linear, a linear model may not be sufficient for all control tasks; however, for regulation and trajectory tracking tasks, 
%it is a common practice to linearize the system dynamics around the reference set-point/trajectory and use the linearized dynamics to design the controller. Thus, for such tasks, 
a linear model is often adequate (see Sec.~\ref{sec:sim}). 
A linear parameterization is also used in adaptive control for similar reasons \cite{sastry2011adaptive}. 
Nevertheless, the proposed framework can handle more general model classes as long as the optimal control problem in \eqref{eqn:lin_ctrl} can be solved for those classes.  

Since $\dyn$ is unknown, the shape of the cost function, $\cost(\param{})$, in \eqref{eqn:optzProb} is unknown. The cost is thus evaluated empirically in each experiment, which is often expensive as it involves conducting an experiment.
%, which may last for a few minutes. 
Thus, the goal is to solve the optimization problem in \eqref{eqn:optzProb} with as few evaluations as possible. 
In this paper, we do so via BO.
% Problem formulation (1.5p)
%
% !TEX root = SysIdLQR.tex
\section{Background} \label{background}
In order to optimize $(\paramA{},\paramB{})$, we use BO. 
In this section, we briefly introduce Gaussian processes and BO.

\subsection{Gaussian Process (GP)} \label{sec:gp:gp}
Since the function $\cost(\param{})$ in \eqref{eqn:optzProb} is unknown a priori, we use nonparametric GP models to approximate it over its domain~$\thetaReg$. 
GPs are a popular choice for probabilistic non-parametric regression, where the goal is to find a nonlinear map, $\cost(\param): \thetaReg \rightarrow \R$, from an input vector $\param \in \thetaReg$ to the function value $\cost(\param)$. 
Hence, we assume that function values $\cost(\param)$, associated with different values of $\param{}$, are random variables and that any finite number of these random variables have a joint Gaussian distribution dependent on the values of $\param{}$~\cite{Rasmussen2006}.
For GPs, we define a prior mean function and a covariance function (or kernel), $k(\parameters_i,\parameters_j)$, which defines the covariance between any two function values, $\cost(\parameters_i)$ and $\cost(\parameters_j)$. 
In this work, the mean is assumed to be zero without loss of generality.
%
%	Gaussian Processes (GPs) are a state-of-the-art probabilistic non-parametric regression method~\citep{Rasmussen2006}. 
%	A GP is a distribution over functions
%	%
%	\begin{align}
%		\respsurfNo \sim \GP \left( m,\coveq \right)\,,
%	\end{align}
%	%
%	fully defined by a mean function $m$ and a {\gpkernel}~$\coveq$. 
%	In this work, the mean is assumed to be zero without loss of generality. 
The choice of kernel is problem-dependent and encodes general assumptions such as smoothness of the unknown function. 
In the experimental section, we employ the $5/2$ Mat\`ern kernel where the hyperparameters are optimized by maximizing the marginal likelihood~\citep{Rasmussen2006}. 
This kernel function implies that the underlying function $\cost$ is differentiable and takes values within the $2\sigma_f$ confidence interval with high probability. %These hyperparameters thus encode our prior assumptions about the unknown performance function. 

The GP framework can be used to predict the distribution of the performance function~$\cost(\param^{*})$ at an arbitrary input~$\param^{*}$ based on the past observations, $\dataset=\{\param_i,\cost(\param_i)\}_{i=1}^n$. 
Conditioned on $\dataset$, the mean and variance of the prediction are
%	%
%	\begin{align}
%		\prob \left( \respsurf{\parameters_*}|\D,\parameters_* \right) & = \gauss{\mu(\parameters_*)}{\sigma^2(\parameters_*)}\,, 
%		\label{eq:one-step prediction distr}\\
%		\mu(\parameters_*) &= \vec k^T_*(\mat K + \sigma_w^2 \mat I)^{-1} \outputVec\,,
%		\label{eq:one-step prediction mean}\\
%		\sigma^2(\parameters_*) &= k_{**}-\vec k^T_*(\mat K + \sigma_w^2 \mat I)^{-1}\vec k_*\,,
%	  	\label{eq:one-step prediction mean and covariance}
%	\end{align}
%	%
	%
\begin{equation} \label{eq:one-step prediction mean and covariance}\\
\mu(\param^{*}) = {\vec k}\mat K^{-1} {\vec \cost};~~ \sigma^2(\param^{*}) = k(\param^{*},\param^{*})-{\vec k}\mat K^{-1}{\vec k}^T\,,
\end{equation}
where $\mat K$ is the kernel matrix with $K_{ij}= k(\parameters_i,\parameters_j)$, $\vec k =[k(\param_1,\param^{*}),\ldots,k(\param_n,\param^{*})]$ and $\vec \cost =[\cost(\param_1),\ldots,\cost(\param_n)]$. Thus, the GP provides both the expected value of the performance function at any arbitrary point~$\param^{*}$ as well as a notion of the uncertainty of this estimate. 

\subsection{Bayesian Optimization (BO)} \label{BO}
Bayesian optimization aims to find the global minimum of an unknown function~\cite{Kushner1964,Osborne2009,Shahriari2016}. 
BO is particularly suitable for the scenarios where evaluating the unknown function is expensive, which fits our problem in Sec. \ref{sec:formulation}.
At each iteration, BO uses the past observations $\dataset$ to model the objective function, and uses this model to determine informative sample locations. 
%This data set is used to build a model~$\respsurf{\cdot}: \parameters \mapsto \objfunc{\parameters}$, called the \textit{response surface}, that maps the parameters $\parameters$ to the corresponding function evaluations~$\objfunc{\parameters}$. 
A common model used in BO for the underlying objective, and the one that we consider, are Gaussian processes (see Sec. \ref{sec:gp:gp}). 
Using the mean and variance predictions of the GP from \eqref{eq:one-step prediction mean and covariance}, BO computes the next sample location by optimizing the so-called acquisition function, $\acqfunc{\cdot}$.
% GP-based methods use the mean and variance predictions in \eqref{eq:one-step prediction mean and covariance} to compute the next sample location by optimizing the so-called acquisition function, $\acqfunc{\cdot}$. 
Different acquisition functions are used in literature to trade off between exploration and exploitation during the optimization process~\cite{Shahriari2016}. 
For example, the next evaluation for expected improvement (EI) acquisition function~\cite{movckus1975bayesian} is given by $\param^{*} = \arg\min_{\param}\acqfunc{\param}$ where 
\begin{equation} \label{eqn:acqfunc}
\acqfunc{\param} = \sigma(\param)[u\Phi(u)+\phi(u)];\quad u= (\mu(\param)-T)/\sigma(\param).
\end{equation}
$\Phi(\cdot)$ and $\phi(\cdot)$ in \eqref{eqn:acqfunc}, respectively, are the standard normal cumulative distribution and probability density functions. 
The target value $T$ is the minimum of all explored data. 
Intuitively, EI selects the next parameter point where the expected improvement over $T$ is maximal. 
Repeatedly evaluating the system at points given by \eqref{eqn:acqfunc} thus improves the observed performance.
Note that optimizing $\acqfunc{\param}$ in \eqref{eqn:acqfunc} does not require physical interactions with the system, but only evaluation of the GP model. 
When a new set of optimal parameters~$\parameters^*$ is determined, they are finally evaluated on the real objective function~$\objfuncNo$ (i.e., the system). 
% Bayesian Optimization and Gaussian Processes
%
% !TEX root = SysIdLQR.tex
\section{Dynamics Optimization via BO (\metName{})} \label{solution}
This section presents the technical details of \metName{}, a novel framework for optimizing dynamics model for maximizing the resultant controller performance. In this work,~$\param \in \R^{n_x(n_x+n_u)}$, i.e., each dimension in $\param$ corresponds to an entry of the $\paramA{}$ or $\paramB{}$ matrices. This parameterization is chosen for simplicity, but other parameterizations can easily be used. 

Given an initial state of the system $\state_0$ and the current system dynamics model $(\paramA{\intsymbol},\paramB{\intsymbol})$, we design an optimal control sequence $\pol_0(\state_0,\param^{\intsymbol})$ that minimizes the cost function $\cost_0(\stateseq_0^N,\ctrlseq_0^{N-1})$, i.e., we solve the optimal control problem in \eqref{eqn:lin_ctrl}. The first control of this control sequence is applied on the \textit{actual system} and the next state $\state_1$ is measured. We then similarly compute $\pol_1(\state_1,\param^{\intsymbol})$ starting at $\state_1$, apply the first control in the obtained control sequence, measure $\state_2$, and so on until we get $\state_N$. Once $\stateseq_0^N$ and $\ctrlseq_0^{N-1}$ are obtained, we compute the true performance of $\ctrlseq_0^{N-1}$ on the actual system by analytically computing $\cost_0(\stateseq_0^N,\ctrlseq_0^{N-1})$ using \eqref{eqn:nl_ctrl}. We denote this cost by $\cost(\param^{\intsymbol})$ for simplicity. We next update the GP based on the collected data sample $\{\param^{\intsymbol},\cost(\param^{\intsymbol})\}$. Finally, we compute $\param^*$ that minimizes the corresponding acquisition function $\acqfunc{\param}$ and repeat the process for $(\paramA{*},\paramB{*})$. Our approach is illustrated in Figure \ref{fig:framework} and summarized in \alg{alg:solution}.
\begin{algorithm}[tb]
	%\RestyleAlgo{boxed}
	\DontPrintSemicolon
	%\AlgoDisplayBlockMarkers\SetAlgoBlockMarkers{}{end}%
	%\SetAlgoNoEnd
	\caption{\metName{} algorithm}
	\label{alg:solution}
	$\dataset$ \hspace{2.9 mm} $\longleftarrow$ if available: $\{\parameters, \objfunc{\parameters}\}$\;
	Prior $\longleftarrow$ if available: Prior of the GP hyperparameters\;	
	Initialize GP with $\dataset$\;	
	\While{\text{optimize}}{	
		Find $\param^* = \arg\min_{\param} \acqfunc{\param}$;\quad $\param^{\intsymbol} \longleftarrow \param^*$ \;	
		$\stateseq_0^N = \{ \}$, $\ctrlseq_0^{N-1} = \{ \}$\;	
		\For{\text{$i=0:N-1$}}{
			Given $\state_i$ and $(\paramA{\intsymbol},\paramB{\intsymbol})$, compute $\pol_i(\state_i,\param^{\intsymbol})$\;
			Apply $\pol_i^1(\state_i,\param^{\intsymbol})$ on the real system and measure $\state_{i+1}$\;
			$\stateseq_0^N \longleftarrow (\stateseq_0^N, \state_{i+1})$\;
			$\ctrlseq_0^{N-1} \longleftarrow (\ctrlseq_0^{N-1},\pol_i^1(\state_i, \param^{\intsymbol}))$\;
		}
		Evaluate $\cost(\param^{\intsymbol}) := J_0(\stateseq_0^N, \ctrlseq_0^{N-1})$ using \eqref{eqn:nl_ctrl}\;
		Update GP and $\dataset$ with $\{\param^{\intsymbol},\cost(\param^{\intsymbol})\}$\;
		}
\end{algorithm}
Intuitively, \metName{} directly learns the shape of the cost function $\cost(\param{})$ as a function of linearizations $(\paramA{},\paramB{})$. Instead of learning the global shape of this function through random queries, it analyzes the performance of all the past evaluations and by optimizing the acquisition function, generates the next query that provides the maximum information about the minima of the cost function. This direct \textit{minima-seeking} behavior based on the \textit{actual observed performance} ensures that our approach is data-efficient. Thus, in the space of all linearizations, we efficiently and directly search for the linearization whose corresponding controller minimizes $J_0$ on the actual system.

Since the problem in \eqref{eqn:lin_ctrl} is an optimal control problem for the linear system $(\paramA{\intsymbol},\paramB{\intsymbol})$, depending on the form of the cost function $\cost$, different optimal control schemes can be used. For example, if $\cost$ is quadratic, the optimal controller is a linear feedback controller given by the solution of a Riccati equation. If $\cost$ is a general convex function, the optimal control problem is solved through a general convex MPC solver, and the resultant controller could be non-linear. Thus, depending on the form of $\cost$, the controller designed by \metName{} can be linear or non-linear. This property causes \metName{} to perform well in the scenarios where a linear controller is not sufficient, as shown in Sec.~\ref{sec:comp2}. More generally, the proposed framework is modular and other control schemes can be used that are more suitable for a given cost function, which allows us to capture a richer controller space. %In theory, for a given cost function, by learning $(A, B)$ we can capture every controller (linear or non-linear) that is an optimal controller for \textit{some} linear system. 

Note that the GP in our algorithm can be initialized with dynamics models whose controllers are known to perform well on the actual system. This generally leads to a faster convergence. For example, when a good linearization of the system is known, it can be used to initialize $\dataset$. When no information is known about the system a priori, the initial models are queried randomly. Finally, note that \metName{} can also be used when the real system is stochastic. In this case, \metName{} will minimize the expected cost.
% Solution methodology
% Overall algorithm
%
% !TEX root = SysIdLQR.tex
\section{Numerical Simulations \label{sec:sim}}
In this section, we present some simulation results on the performance of the proposed method for controller design. 

\subsection{Dubins Car System \label{sec:dubin_sim}} 
\begin{figure}[t]
  \centering
  \includegraphics[width=0.4\textwidth]{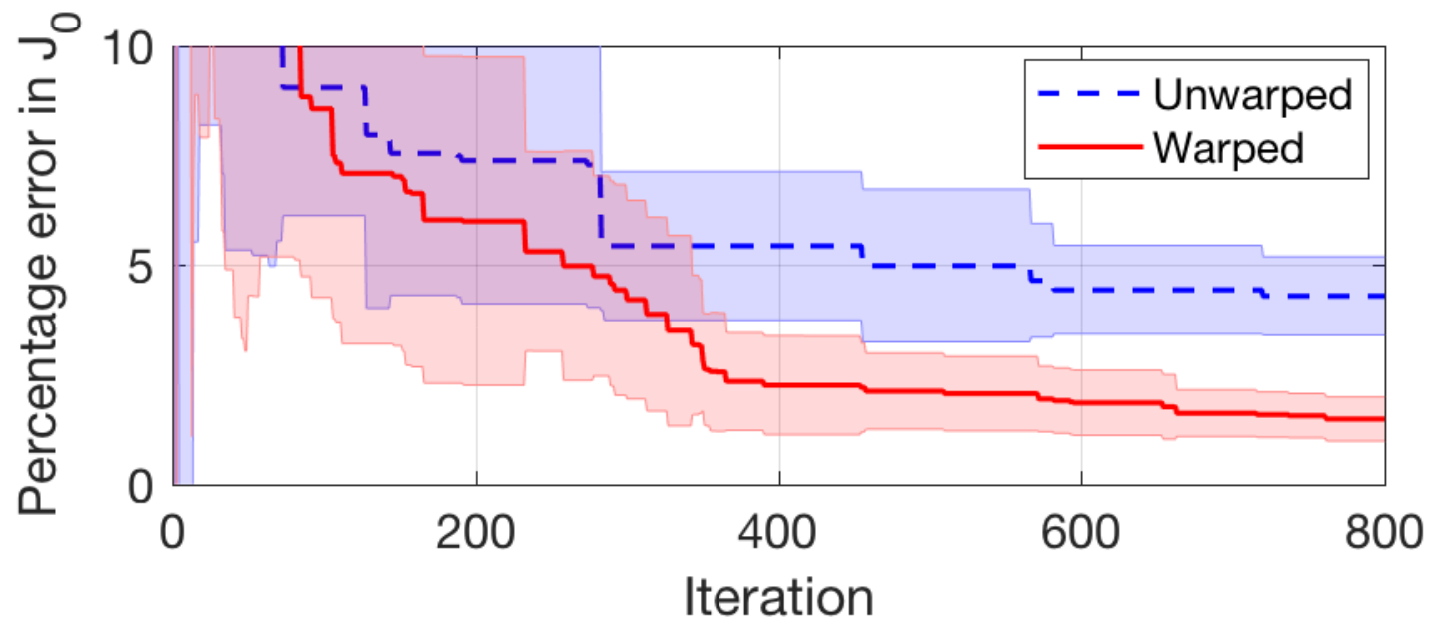}
  \caption{Dubins car: mean and standard deviation of $\compMet$ during the learning process (over 10 trials). 
  \metName{} reaches within the 10\% of the optimal cost in just 100 iterations, starting from a random dynamics model.
  Using a log warping on the cost function further accelerates the learning.
 % Using the log warping the learned controller reaches within 6\% of the optimal cost in 200 iterations, outperforming the unwarped case.
 }
  \label{fig:NumSim_fig1}
\end{figure}
For the first simulation, we consider a three dimensional non-linear Dubins car whose dynamics are given as
%%
%\begin{equation} \label{eqn:NumSimpleDyn}
%\begin{aligned} 
%x_{k+1} &=  x_k + Tv_k \cos\theta_k\,, \\
%y_{k+1} &= y_k + Tv_k \sin\theta_k\,, \\
%\theta_{k+1} &= \theta_k + T\omega_k\,,
%\end{aligned}
%\end{equation}
%%
%where $\state_k := (x_k, y_k, \theta_k)$ is the state of system, $\pos_k = (x_k, y_k)$ is the position, $\theta_k$ is the heading, $v_k$ is the speed, and $\omega_k$ is the turn rate at time $k$. The input (control) to the system is $\ctrl_k := (v_k, \omega_k)$. $T$ represents the sampling time and is chosen to be $0.1$s in our simulations. 
%
\begin{equation} \label{eqn:NumSimpleDyn}
\dot{x} = v\cos\phi,\quad \dot{y} = v\sin\phi,\quad \dot{\phi} = \omega\,,
\end{equation}
where $\state := (x, y, \phi)$ is the state of system, $\pos = (x, y)$ is the position, $\phi$ is the heading, $v$ is the speed, and $\omega$ is the turn rate. The input (control) to the system is $\ctrl := (v, \omega)$. For simulation purposes, we discretize the dynamics at a frequency of 10Hz. %The initial state of the system is chosen as $\state_0 := (1.5, 1, \pi/2)$. 
Our goal is to design a controller that steers the system to the equilibrium point $\state^* = 0, \ctrl^* = 0$ starting from the state $\state_0 := (1.5, 1, \pi/2)$. In particular, we want to minimize the cost function
\begin{equation} \label{eqn:costSim}
\cost_0(\stateseq_0^N, \ctrlseq_0^{N-1})  =\sum_{k=0}^{N-1}\left( \state_k^T Q \state_k + \ctrl_k^T R \ctrl_k \right) + \state_N^T Q_f \state_N\,.
\end{equation}
We choose $N = 30$. $Q$, $Q_f$ and $R$ are all chosen as identity matrices of appropriate sizes. 
We also assume that the dynamics are not known; hence, we cannot directly design a controller to steer the system to the desired equilibrium. 
Instead, we use \metName{} to find a linearization of dynamics in \eqref{eqn:NumSimpleDyn} that minimizes the cost function in \eqref{eqn:costSim}, directly from the experimental data. 
In particular, we represent the system in \eqref{eqn:NumSimpleDyn} by a parameterized linear system $\state_{k+1} = \paramA{}\state_{k} + \paramB{}\ctrl_{k}$, design a controller for this system and apply it on the actual system. Based on the observed performance, BO suggests a new linearization and the process is repeated.
Since the cost function is quadratic in this case, the optimal control problem for a particular $\param{}$ is an LQR problem, and can be solved efficiently. 
%The resulting controller is applied in closed-loop on the actual system in \eqref{eqn:NumSimpleDyn} and performance is recorded. 
%
%
\begin{figure}[t]
  \centering
  \includegraphics[width=0.4\textwidth]{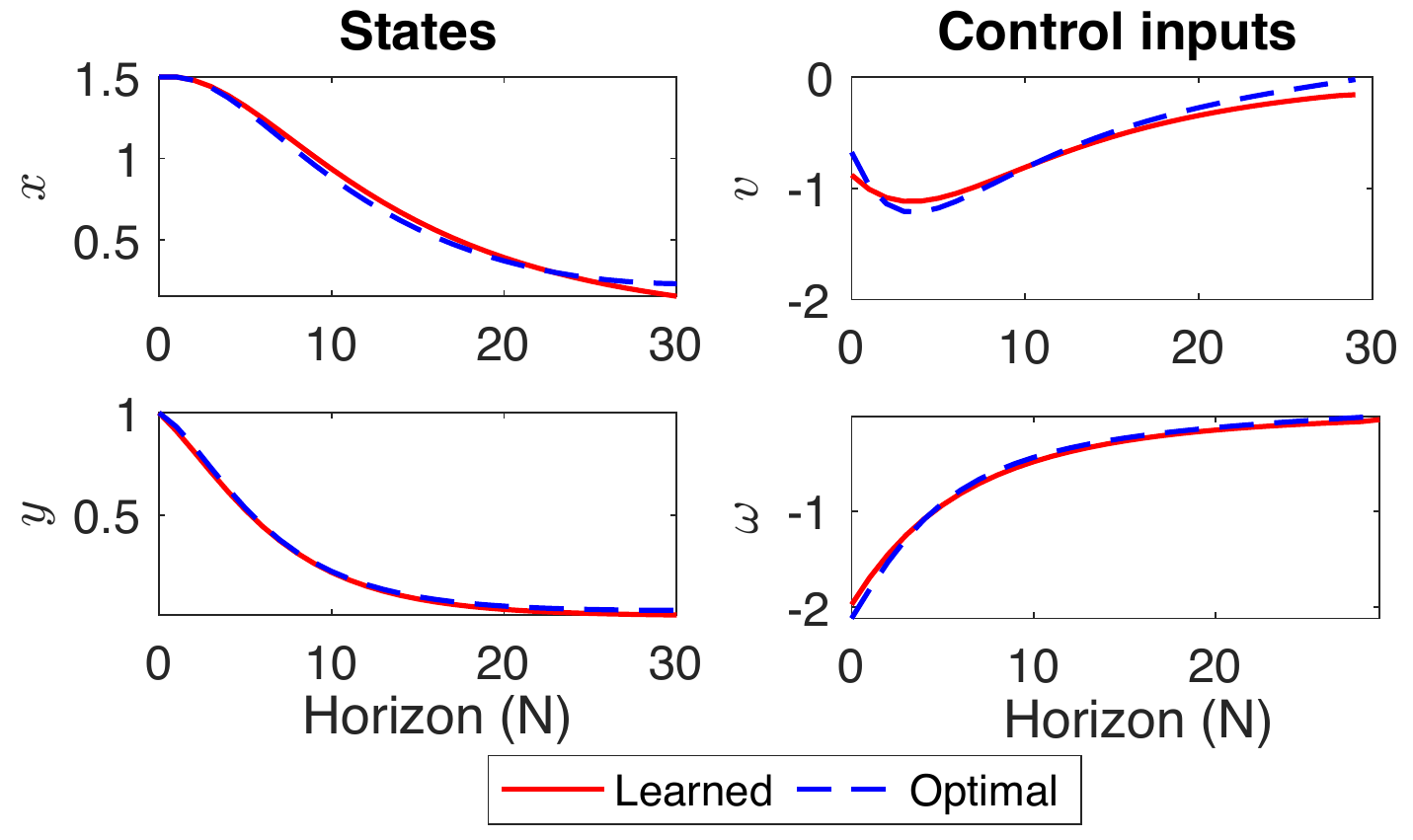}
  \caption{Dubins car: state and control trajectories for the learned and the true system. The two trajectories are very similar, indicating that the learned dynamics model represents the system behavior accurately around the desired state.}
  \label{fig:NumSim_statecontrol}
\end{figure}
For BO, we use the MATLAB library \textit{BayesOpt}~\cite{martinez2014bayesopt}. 
Since there are $3$ states and $2$ inputs, we learn $15$ parameters in total, one corresponding to each entry of the $\paramA{}$ and $\paramB{}$ matrices. 
The bounds on the parameters are chosen randomly as $\thetaReg = [-2, 2]^{15}$. 
As acquisition function, we use EI (see eq. \eqref{eqn:acqfunc}). 
Since no information is assumed to be known about the system, the GP was initialized with a random $\param{}$.
We also warp the cost function $\cost$ using the $log$ function before passing it to BO. 
Warping makes the cost function smoother while maintaining its monotonic properties, which makes the sampling process in BO more efficient and leads to a faster convergence. 

For comparison, we compute the true optimal controller that minimizes \eqref{eqn:costSim} subject to the dynamics in \eqref{eqn:NumSimpleDyn} using the non-linear solver \textit{fmincon} in MATLAB to get the minimum achievable cost $\cost_0^*$ across all controllers. 
We use the percentage error between the true optimal cost $\cost_0^*$ and the cost achieved by \metName{} as our comparison metric in this work
\begin{equation} \label{eqn:comp_metric}
\compMet_n = 100 \times (\cost_0^* - \cost(\param_n))/\cost_0^*\,,
\end{equation}
where $\cost(\param_n)$ is the best cost achieved by aDOBO by iteration~$n$. 
In Fig. \ref{fig:NumSim_fig1}, we plot $\compMet_n$ for Dubins car. As learning progresses, \metName{} gathers more and more information about the minimum of $\cost_0$ and reaches within 10\% of $\cost_0^*$ in just 100 iterations, demonstrating its effectiveness in designing a controller for an unknown system just from the experimental data. Fig. \ref{fig:NumSim_fig1} also highlights the effect of warping in BO. A well warped function converges faster to the optimal performance.
We also compared the control and state trajectories obtained from the learned controller with the optimal control and state trajectories. As shown in Fig. \ref{fig:NumSim_statecontrol}, the learned system matrices not only achieve the optimal cost, but also follow the optimal state and control trajectories very closely. 
Even though the trajectories are very close to each other for the true system and its learned linearization, this linearization may not correspond to any \textit{actual} linearization of the system. %In other words, even if the true system was linear, the learned system matrices may not be same as the actual system matrices. 
The next simulation illustrates this key property of \metName{} more clearly.  

\subsection{A Simple 1D Linear System \label{sec:simple1D}} 
For this simulation, we consider a simple 1D linear system
\begin{equation} \label{eqn:simpleLinDyn}
\state_{k+1} =  \state_k + \ctrl_k\,,
\end{equation}
where $\state_k$ and $\ctrl_k$ are the state and the input of the system at time $k$. Although the dynamics model is very simple, it illustrates some key insights about the proposed method. %The initial state of the system is chosen as $\state_0 = 1$. 
Our goal is to design a controller that minimizes \eqref{eqn:costSim} starting from the state $\state_0 = 1$. We choose $N = 30$ and $R = Q = Q_f =1$. We assume that the dynamics are unknown and use \metName{} to learn the dynamics, where $\param{} := (\param_1, \param_2) \in \R^2$ are the dynamics parameters to be learned.

\begin{wrapfigure}{r}{0.5\columnwidth}
%  \begin{figure}[t]
  \centering
  \vspace{-15pt}
  \includegraphics[width=0.48\columnwidth]{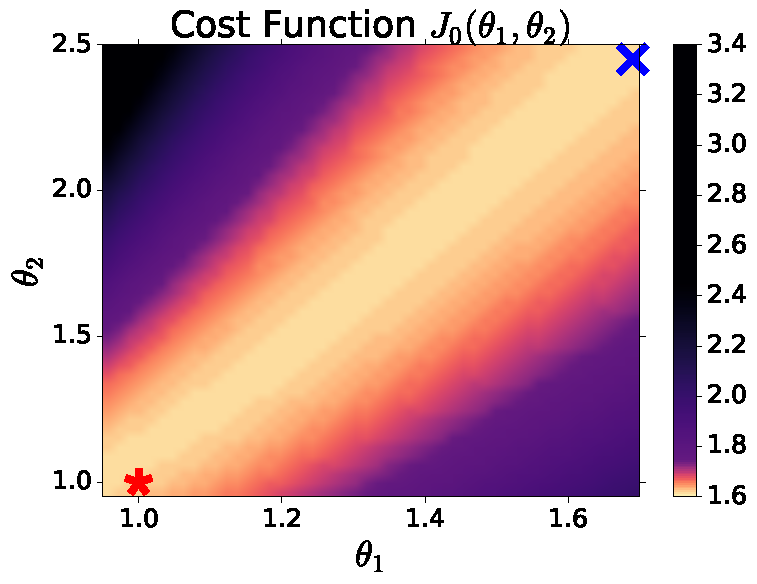}
  \caption{Cost of the actual system in \eqref{eqn:simpleLinDyn} as a function of the linearization parameters $(\param_1, \param_2)$. 
  The parameters obtained by \metName{} (the pink X) yield to performance very close to the true system parameters (the green $*$).
  Note that \metName{} does not necessarily converge to the true parameters.
%   The true system parameters and the converged parameters are denoted by the Green $*$ and the Pink $X$ in the figure. 
%   The two sets of parameters yield very close performance on the actual system, but  has relatively the same performance for , causing \metName{} to not necessarily converge to the true parameters.
  }
  \label{fig:Jvisual}
  \vspace{-5pt}
% \end{figure}
\end{wrapfigure}
The learning process converges in $45$ iterations to the true optimal performance ($\cost_0^* = 1.61$), which is computed using LQR on the real system. 
The converged parameters are $\param_1 = 1.69$ and $\param_2 = 2.45$, which are vastly different from the true parameters $\param_1 = 1$ and $\param_2 = 1$, even though the actual system is a linear system. To understand this, we plot the cost obtained on the true system $\cost_0$ as a function of linearization parameters $(\param_1, \param_2)$ in Fig.~\ref{fig:Jvisual}. Since the performances of the two sets of parameters are very close to each other, a direct performance based learning process (e.g., \metName{}) cannot distinguish between them and both sets are equally optimal for it. 
More generally, a wide range of parameters lead to similar performance on the actual system. Hence, we can expect the proposed approach to recover the optimal controller and the actual state trajectories, but not necessarily the true dynamics or its true linearization. This simulation also suggests that the true dynamics of the system may not even be required as far as the control performance is concerned.

\subsection{Cart-pole System \label{sec:cartpole}}
We next apply \metName{} to a cart-pole system
\begin{equation} \label{eqn:cartpole_dyn}
\begin{aligned}
(M+m)\ddot{x} & - ml\ddot{\psi}\cos\psi + ml{\dot{\psi}}^2\sin\psi = F\,,\\
l\ddot{\psi} - g\sin\psi & = \ddot{x} \cos\psi\,,
\end{aligned}
\end{equation}
where $x$ denotes the position of the cart with mass $M$, $\psi$ denotes the pendulum angle, and $F$ is a force that serves as the control input. 
The massless pendulum is of length $l$ with a mass $m$ attached at its end. Define the system state as $\state := (x, \dot{x}, \psi, \dot{\psi})$ and the input as $\ctrl := F$. 
Starting from the state $(0, 0, \tfrac{\pi}{6}, 0)$, the goal is to keep the pendulum straight up, while keeping the state within given lower and upper bounds. 
In particular, we want to minimize the cost 
\begin{equation} \label{eqn:costMPC}
\begin{aligned}
J_0(\stateseq_0^N, \ctrlseq_0^{N-1}) = & \sum_{k=0}^{N-1}\left( \state_k^T Q \state_k + \ctrl_k^T R \ctrl_k \right) + \state_N^T Q_f \state_N \\
& + \lambda\sum_{i=0}^{N} \max(0, \underline{\state} - \state_i, \state_i - \overline{\state}),
\end{aligned}
\end{equation}
where $\lambda$ penalizes the deviation of state $\state_i$ below $\underline{\state}$ and above $\overline{\state}$. We assume that the dynamics are unknown and use \metName{} to optimize the dynamics. 
%
% %
% \begin{figure}[t]
% \centering
% \begin{subfigure}[b]{.4\columnwidth}
%   \centering
%   \includegraphics[height=2.8cm]{"figs/cartpole_sys"}
%   \subcaption{Cart-pole System}
%   \label{fig:cartpole_sys}
% \end{subfigure}%
% \hfill
% \begin{subfigure}[b]{.58\columnwidth}
% \centering
%   \includegraphics[height=2.8cm]{"figs/crazyflie_pic"}
%   \subcaption{Crazyflie 2.0}
%   \label{fig:crazyflie}
% \end{subfigure}
% \caption{The Cart-pole and the Crazyflie systems}
% \label{fig:systems}
% \end{figure}
% %
%
For simulation, we discretize the dynamics at a frequency of $10$Hz. We choose $N=30$, $M=1.5$Kg, $m=0.175$Kg, $\lambda = 100$ and $l=0.28$m. The $Q=Q_f=\diag([0.1, 1, 100, 1])$ and $R=0.1$ matrices are chosen to penalize the angular deviation significantly. We use $\underline{\state}=[-2, -\infty, -0.1, -\infty]$ and $\overline{\state}=[2, \infty, \infty, \infty]$, i.e., we are interested in controlling the pendulum while keeping the cart position within $[-2, 2]$, and limiting the pendulum overshoot to $0.1$. The optimal control problem for a particular linearization is a convex MPC problem and solved using YALMIP \cite{lofberg2005yalmip}. The true $\cost_0^*$ is computed using \textit{fmincon}. 

\begin{figure}[t]
  \centering
  \includegraphics[width=0.75\columnwidth]{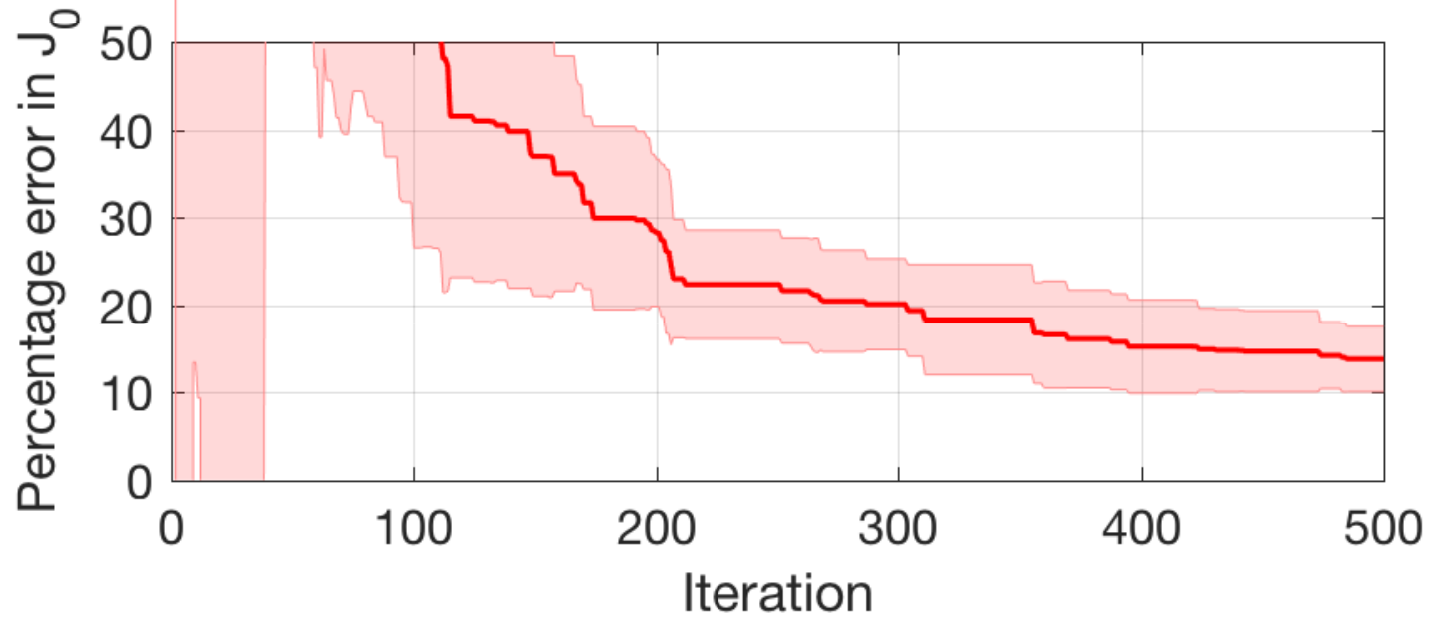}
  \caption{Cart-pole system: mean and standard deviation of $\compMet$ during the learning process. The learned controller reaches within 20\% of the optimal cost in 250 iterations, demonstrating the applicability of \metName{} to highly non-linear systems.}
  \label{fig:NumSim_fig2}
\end{figure}

As shown in Fig. \ref{fig:NumSim_fig2}, \metName{} reaches within 20\% of the optimal performance in 250 iterations and continue to make progress towards finding the optimal controller. This simulation demonstrates that the proposed method (a) is also applicable to highly non-linear systems, (b) can handle general convex cost functions that are not necessarily quadratic, and (c) different optimal control schemes can be used within the proposed framework. Since an MPC controller can in general be non-linear, this implies that the proposed method can also design complex non-linear controllers with an LTI parametrization. 
% Numerical Simulations 
%
% !TEX root = SysIdLQR.tex
\section{Comparison with other methods \label{sec:comparison}}
In this section, we compare our approach with some other online learning schemes for controller design.%, and discuss their relative advantages and disadvantages.  

\subsection{Tuning $(Q,R)$ vs \metName{}} \label{sec:comp1}
In this section, we consider the case in which the cost function $\cost_0$ is quadratic (see Eq. \eqref{eqn:costSim}). Suppose that the actual linearization of the system around $\state^*= 0$ and $\ctrl^*= 0$ is known and given by $(A^*, B^*)$. To design a controller for the actual system in such a case, it is a common practice to use an LQR controller for the linearized dynamics. However, the resultant controller may be sub-optimal for the actual non-linear system. To overcome this problem, authors in \cite{Marco2016, trimpe2014self} propose to optimize the controller by tuning penalty matrices $Q$ and $R$ in \eqref{eqn:costSim}. 
In particular, we solve
\begin{equation} \label{eqn:optzProb_QR}
\begin{aligned} 
\theta^* &= \arg \min_{\theta \in \thetaReg} J_0(\stateseq_0^N, \ctrlseq_0^{N-1})\,, \\
\text{sub. to}~~ \state_{k+1} &= \dyn(\state_k, \ctrl_k),\quad \ctrl_k = K(\theta)\state_k\,,\\
K(\theta) &= LQR(A^*, B^*, W_Q(\theta), W_R(\theta), Q_f)\,,
\end{aligned} 
\end{equation}
where $K(\theta)$ denotes the LQR feedback matrix obtained for the system matrices $(A^*,B^*)$ with $W_Q$ and $W_R$ as state and input penalty matrices, and can be computed analytically. For further details of LQR method, we refer interested readers to \cite{bender1987linear}. The difference between optimization problems \eqref{eqn:optzProb} and \eqref{eqn:optzProb_QR} is that now we parameterize penalty matrices $W_Q$ and $W_R$ instead of system dynamics. %, and tune them to achieve the optimal performance on the actual non-linear system directly using the experimental data. 
The optimization problem in \eqref{eqn:optzProb_QR} is solved using BO in a similar fashion as we solve \eqref{eqn:optzProb} \cite{Marco2016}. 
The parameter~$\theta$, in this case, can be initialized by the actual penalty matrices $Q$ and $R$, instead of a random query, which generally leads to a much faster convergence.
% as the true linearization is generally not too far from the optimal performance. 
An alternative approach is to use \metName{}, except that now we can use $(A^*,B^*)$ as initializations for the system matrices $A$ and $B$. 
Actual penalty matrices $Q$ and $R$ are used for \metName{}. %Thus, in this scenario and in general when a good linearization is available, the proposed method can be thought of as a controller tuning method rather than controller design method.   

\begin{figure}[t]
  \centering
  \includegraphics[width=0.75\columnwidth]{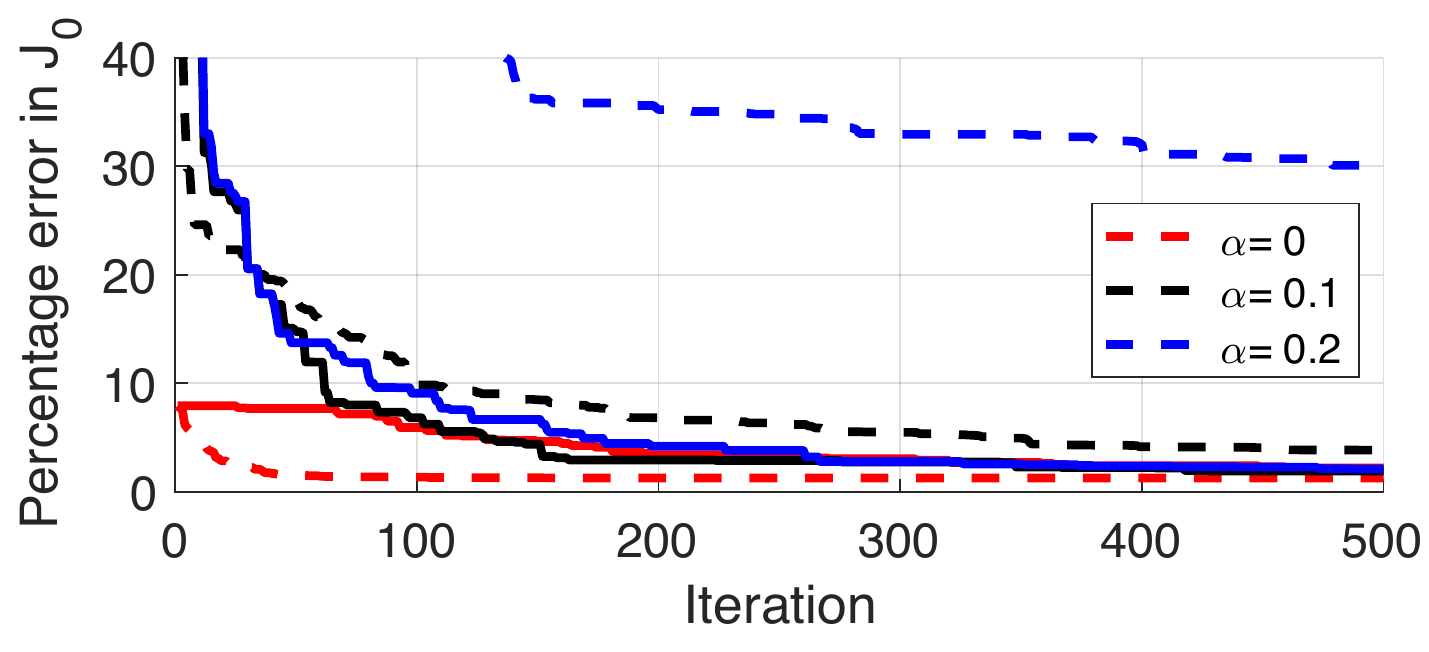}
  \caption{Dubins car: Comparison between tuning the penalty matrices $(Q,R)$ \cite{Marco2016} (dashed curves), and \metName{} (solid curves) for different noise levels in $(A^*,B^*)$, the true linearized dynamics around the desired goal state. When the true linearized dynamics are known perfectly, the $(Q,R)$ tuning method outperforms \metName{} because fewer parameters are to be learned. Its performance, however, drops significantly as noise increases, rendering the method impractical for the scenarios where system dynamics are not known to a good accuracy.}
  \label{fig:QRvsAB_tuning}
\end{figure}

When $(A^*,B^*)$ are known to a good accuracy, $(Q,R)$ tuning method is expected to converge quickly to the optimal performance compared to \metName{} as it needs to learn fewer parameters, i.e., ($n_x+n_u$) %in the case of $(Q,R)$ tuning 
(assuming diagonal penalty matrices) compared to $n_x(n_x+n_u)$ parameters for \metName{}. However, when there is error in $(A^*,B^*)$ (or more generally if dynamics are unknown), the performance of the $(Q,R)$ tuning method can degrade significantly as it relies on an accurate linearization of the system dynamics, rendering the method impractical for control design purposes. To compare the two methods we use the Dubins car model in Eq. \eqref{eqn:NumSimpleDyn}. The rest of the simulation parameters are same as Section \ref{sec:dubin_sim}. We compute the linearization of Dubins car around $\state^* = 0$ and $\ctrl^* = 0$ using \eqref{eqn:NumSimpleDyn} and add random matrices $(A_r,B_r)$ to them to generate $A' = (1-\alpha)A^* + {\alpha}A_r$ and $B' = (1-\alpha)B^* + {\alpha}B_r$. We then initialize both methods with $(A',B')$ for different $\alpha$s. As shown in Fig. \ref{fig:QRvsAB_tuning}, the $(Q,R)$ tuning method outperforms \metName{}, when there is no noise in $(A^*,B^*)$. But as $\alpha$ increases, its performance deteriorates significantly. 
In contrast, \metName{} is fairly indifferent to the noise level, as it does not assume any prior knowledge of system dynamics. 
The only information assumed to be known is penalty matrices $(Q, R)$, which are generally designed by the user and hence are known a priori. 
%The another limitation of tuning $(Q, R)$ is that, by design, it can only be used for a quadratic cost function $\cost_0$, whereas \metName{} can be used for more general cost functions as shown in Sec.~\ref{sec:cartpole}.
%

\subsection{Learning $K$ vs \metName{}} \label{sec:comp2}
\begin{figure*}[t]
\centering
\begin{subfigure}[b]{0.98\columnwidth}
\centering
  \includegraphics[width=0.75\columnwidth]{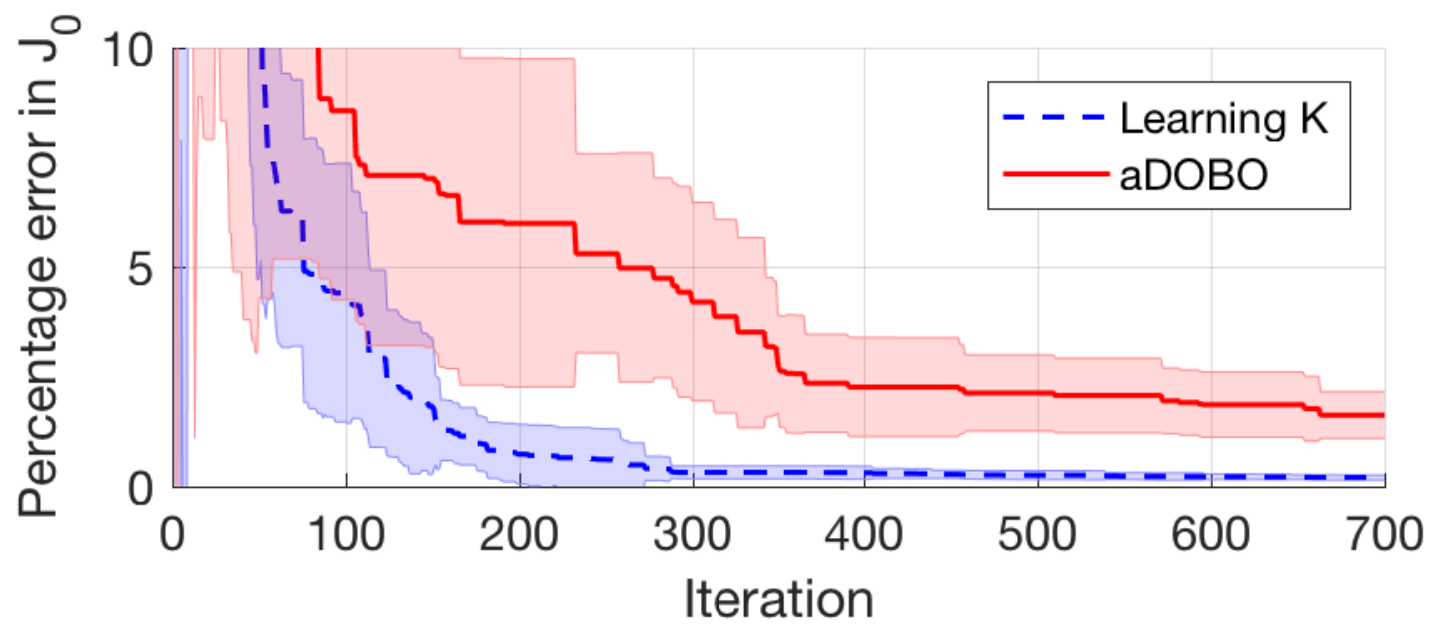}
  \subcaption{Dubins car}
  \label{fig:KvsAB_learning}
\end{subfigure}%
\hfill
\begin{subfigure}[b]{0.98\columnwidth}
\centering
  \includegraphics[width=0.75\columnwidth]{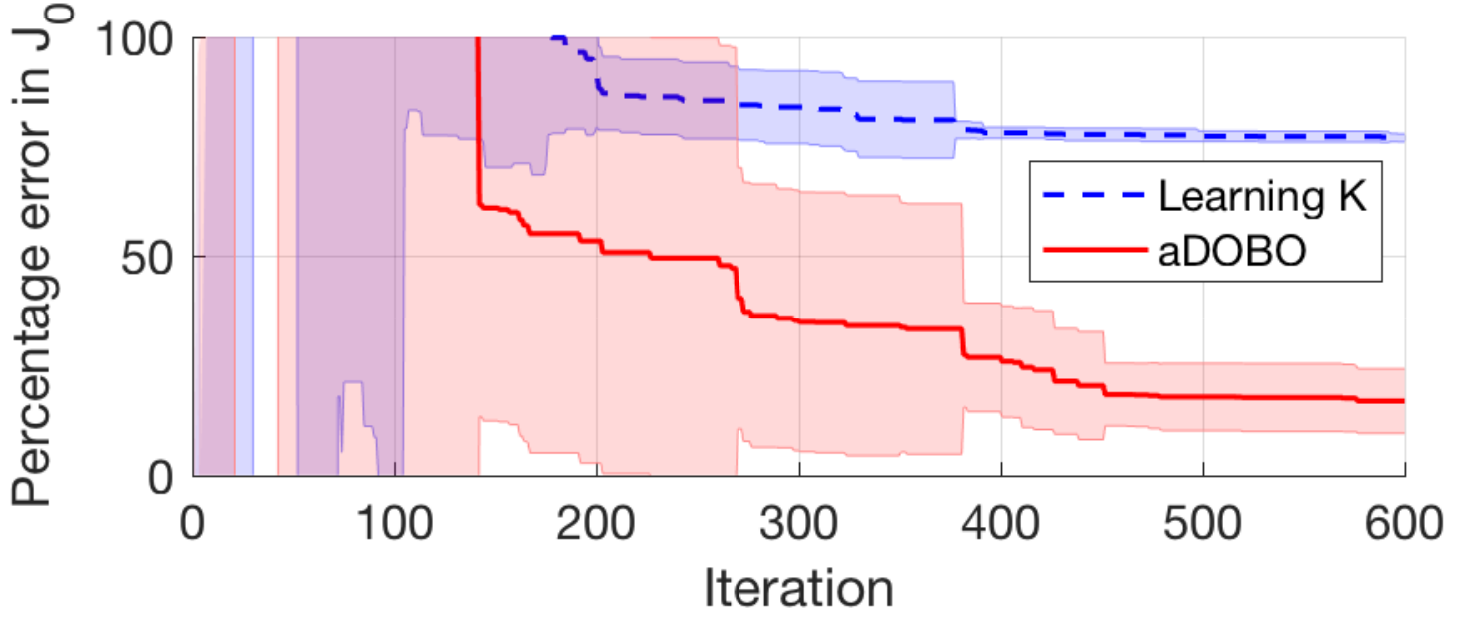}
  \subcaption{System of Eq.~\eqref{eqn:linsys_sim}}
  \label{fig:KvsAB_learning2}
\end{subfigure}
\caption{Mean and standard deviation of $\compMet$ obtained via directly learning the feedback controller $K$ \cite{Calandra2015a} and \metName{} for different cost functions. (a) Comparison for the quadratic cost function of Eq.~\eqref{eqn:costSim}. 
  Directly learning $K$ converges to the optimal performance faster because fewer parameters are to be learned. (b) Comparison for the non-quadratic cost function of Eq.~\eqref{eqn:costMPC}.
  Since the optimal controller for the actual system is not necessarily linear in this case, directly learning $K$ leads to a poor performance}
\label{fig:systems}
\vspace{-7pt}
\end{figure*}

When %system dynamics are unknown and 
the cost function is quadratic, another potential approach is to directly parameterize and optimize the feedback matrix $K \in \R^{n_x n_u}$ in \eqref{eqn:optzProb_QR} \cite{Calandra2015a} as
%%
%\begin{equation} \label{eqn:optzProb_F}
%\begin{aligned}
%\theta^* & =  \arg \min_{\theta \in \thetaReg} J_0(\stateseq_0^N, \ctrlseq_0^{N-1})\,, \\
%\text{sub. to}~~ \state_{k+1} & = \dyn(\state_k, \ctrl_k)\,, \\
%\ctrl_k &= K(\theta)\state_k, K(\theta) \in \mathbb{R}^{n_x \times n_u}\,.
%\end{aligned}
%\end{equation} 
%%
%
\begin{equation} \label{eqn:optzProb_K}
\begin{aligned}
\theta^* & =  \arg \min_{\theta \in \thetaReg} J_0(\stateseq_0^N, \ctrlseq_0^{N-1})\,, \\
\text{sub. to}~~ \state_{k+1} & = \dyn(\state_k, \ctrl_k),\quad \ctrl_k = K_{\theta}\state_k\,.
\end{aligned}
\end{equation} 
The advantage of this approach is that only $n_xn_u$ parameters are learned compared to $n_x(n_x+n_u)$ parameters in \metName{}, which is also evident from Fig.~\ref{fig:KvsAB_learning}, wherein the learning process for $K$ converges much faster than that for \metName{}.
However, a linear controller might not be sufficient for general cost functions, and non-linear controllers are required to achieve a desired performance. As shown in Sec. \ref{sec:cartpole}, \metName{} is not limited to linear controllers; hence, it outperforms the $K$ learning method in such scenarios. Consider, for example, the linear system
%%
%\begin{equation} \label{eqn:linsys_sim}
%\state_{k+1} = \begin{bmatrix} 1 & 1 \\ 0 & 1 \end{bmatrix} \state_k +  \begin{bmatrix} 0 \\ 1 \end{bmatrix} \ctrl_k,
%\end{equation}
%%
%
\begin{equation} \label{eqn:linsys_sim}
x_{k+1} = x_k + y_k,\quad y_{k+1} = y_k + u_k\,,
\end{equation}
and the cost function in Eq. \eqref{eqn:costMPC} with state $\state_k = (x_k, y_k)$, $N = 30$, $\underline{\state} = [0.5, -0.4]$ and $\overline{\state} = [\infty, \infty]$. $Q$, $Q_f$ and $R$ are all identity matrices of appropriate sizes, and $\lambda = 100$. 

As evident from Fig. \ref{fig:KvsAB_learning2}, directly learning a feedback matrix performs poorly with an error as high as 80\% from the optimal cost. Since the cost is not quadratic, the optimal controller is not necessarily linear; however, since the controller in \eqref{eqn:optzProb_K} is restricted to a linear space, it performs rather poorly in this case. In contrast, \metName{} continues to improve performance and reaches within 20\% of the optimal cost within few iterations, because we implicitly parameterize a much richer controller space via learning $A$ and $B$. In this example, we capture non-linear controllers by using a linear dynamics model with a convex MPC solver. 
Since the underlying system is linear, the true optimal controller is also in our search space. Our algorithm makes sure that we make a steady progress towards finding that controller. 
However, we are not restricted to learning a linear controller~$K$. 
One can also directly learn the actual control sequence to be applied to the system (which also captures the optimal controller).
This approach may not be data-efficient compared to \metName{} as the control sequence space can be very large depending on the problem horizon, and will require a large number of experiments. 
As shown in Table~\ref{tab:CtrlvsAB_learning}, the performance error is more than 250\% even after 600 iterations, rendering the method impractical for real systems.   
\begin{table}[t]
\centering
    \begin{tabular}{| c | c | c |}
    \hline
    Iteration & \metName{} & Learning Control Sequence \\ \hline
    200 & \textbf{53 $\pm$ 50\%}  & 605 $\pm$ 420\%  \\ \hline
    400 & \textbf{27 $\pm$ 12\%}  & 357 $\pm$ 159\%   \\ \hline
    600 & \textbf{17 $\pm$ 7\%}  & 263 $\pm$ 150\%  \\
    \hline
    \end{tabular}
    \caption{System in \eqref{eqn:linsys_sim}: mean and standard deviation of $\compMet$ for \metName{}, and for directly learning the control sequence. Since the space of control sequence is huge, the error is substantial even after 600 iterations.}
    \label{tab:CtrlvsAB_learning}  
    \vspace{-5pt}
\end{table}

\subsection{Adaptive Control vs \metName{}} \label{sec:comp3}
In this work, we aim to directly find the best linearization based on the observed performance. Another approach is to learn a true linearization of the system based on the observed state and input trajectory during the experiments. 
The underlying hypothesis is that as more and more data is collected, a better linearization is obtained, eventually leading to an improved control performance. 
This approach is in-line with the traditional model identification and the adaptive control frameworks.  
%
%We record the state and input trajectories in each experiment. 
Let $({}_j\stateseq_0^N, {}_j\ctrlseq_0^{N-1})$ denotes the state and input trajectories for experiment $j$. We also let $\mathcal{D}_i = \cup_{j=1}^i ({}_j\stateseq_0^N, {}_j\ctrlseq_0^{N-1})$. After experiment $i$, we fit an LTI model of the form $\state_{k+1} = A_i \state_k +  B_i \ctrl_k$ using least squares on data in $\mathcal{D}_i$ and then use this model to obtain a new controller for experiment $i+1$. We apply the approach on the linear system in \eqref{eqn:linsys_sim} and the non-linear system in \eqref{eqn:NumSimpleDyn} with the cost function in \eqref {eqn:costSim}. 
For the linear system, the approach converges to the true system dynamics in 5 iterations. However, this approach performs rather poorly on the non-linear system, as shown in Table \ref{tab:adaptCtrl_nl}. 
When the underlying system is non-linear, all state and input trajectories may not contribute to the performance improvement. A good linearization should be obtained from the state and input trajectories in the region of interest, which depends on the task. For example, if we want to regulate the system to the equilibrium $(0,0)$, a linearization of the system around $(0,0)$ should be obtained. Thus, it is desirable to use the system trajectories that are close to this equilibrium point. However, a naive prediction error based approach has no means to select these ``good" trajectories from the pool of trajectories and hence can lead to a poor performance. In contrast, \metName{} does not suffer from these limitations, as it explicitly utilizes a performance based optimization. 
%The result for the non-linear system are shown in Figure \ref{fig:adaptCtrl_nl}.
%%
%\begin{figure}[t]
%  \centering
%  \includegraphics[width=0.4\textwidth]{"figs/adaptCtrl_nl"}
%  \caption{Performance obtained via learning $(A,B)$ through state and input trajectories. When the underlying system is non-linear, this approach can result in a poor performance.}
%  \label{fig:adaptCtrl_nl}
%\end{figure}
%%
%
\begin{table}[t]
\centering
    \begin{tabular}{| c | c | c |}
    \hline
    Iteration & \metName{} & Learning via LS \\ \hline
    200 & \textbf{6 $\pm$ 3.7\%}  & 166.7 $\pm$ 411\%  \\ \hline
    400 & \textbf{2.2 $\pm$ 1.1\%}  & 75.9 $\pm$ 189\%   \\ \hline
    600 & \textbf{1.8 $\pm$ 0.7\%}  & 70.7 $\pm$ 166\%  \\
    \hline
    \end{tabular}
    \caption{Dubins car: mean and standard deviation of $\compMet$ obtained via learning $(A,B)$ through least squares (LS), and through \metName{}.}
    \label{tab:adaptCtrl_nl}  
     \vspace{-3pt}
\end{table}
A summary of the advantages and limitations of the four methods is provided in \tab{table:comparison}.
\begin{table*}
   % \begin{tabular}{ |p{2cm}| p{6cm} | p{\columnwidth} |}
    \begin{tabular}{ |p{3cm}| p{6cm} | p{7.5cm} |}
    \hline
    \textbf{Method} & \textbf{Advantages} & \textbf{Limitations} \\ \hline
    $(Q,R)$ learning \cite{Marco2016} & Only $(n_x+n_u)$ parameters are to be learned so learning will be faster. & Performance can degrade significantly if the dynamics are not known to a good accuracy.\\ \hline
    $F$ learning \cite{Calandra2015a} & Only $n_x n_u$ parameters are to be learned so learning will be faster.  & Approach may not perform well for non-quadratic cost functions. \\ \hline
    $(A,B)$ learning via least squares & Can lead to a faster convergence when the underlying system is linear & Approach is not suitable for non-linear system. \\ \hline
    \metName{} & Does not require any prior knowledge of system dynamics. Applicable to general cost functions. & Number of parameters to be learned is higher, i.e., $(n_x^2+n_x n_u)$. \\
    \hline
    \end{tabular}
    \caption{Relative advantages and limitations of different methods for automatic controller design.}
    \label{table:comparison}
    \squeezeup{-4mm}
\end{table*}
%
% comparison with other similar control design methods
%
% !TEX root = SysIdLQR.tex
\section{Quadrotor Position Tracking Experiments \label{sec:exp}}
%
% \begin{figure}[t]
% \centering
% \begin{subfigure}[b]{.4\columnwidth}
%   \centering
%   \includegraphics[height=2.8cm]{"figs/cartpole_sys"}
%   \subcaption{Cart-pole System}
%   \label{fig:cartpole_sys}
% \end{subfigure}%
% \hfill
% \begin{subfigure}[b]{.58\columnwidth}
\begin{wrapfigure}{r}{0.46\columnwidth}
  \centering
  \vspace{-10pt}
  \includegraphics[width=0.44\columnwidth]{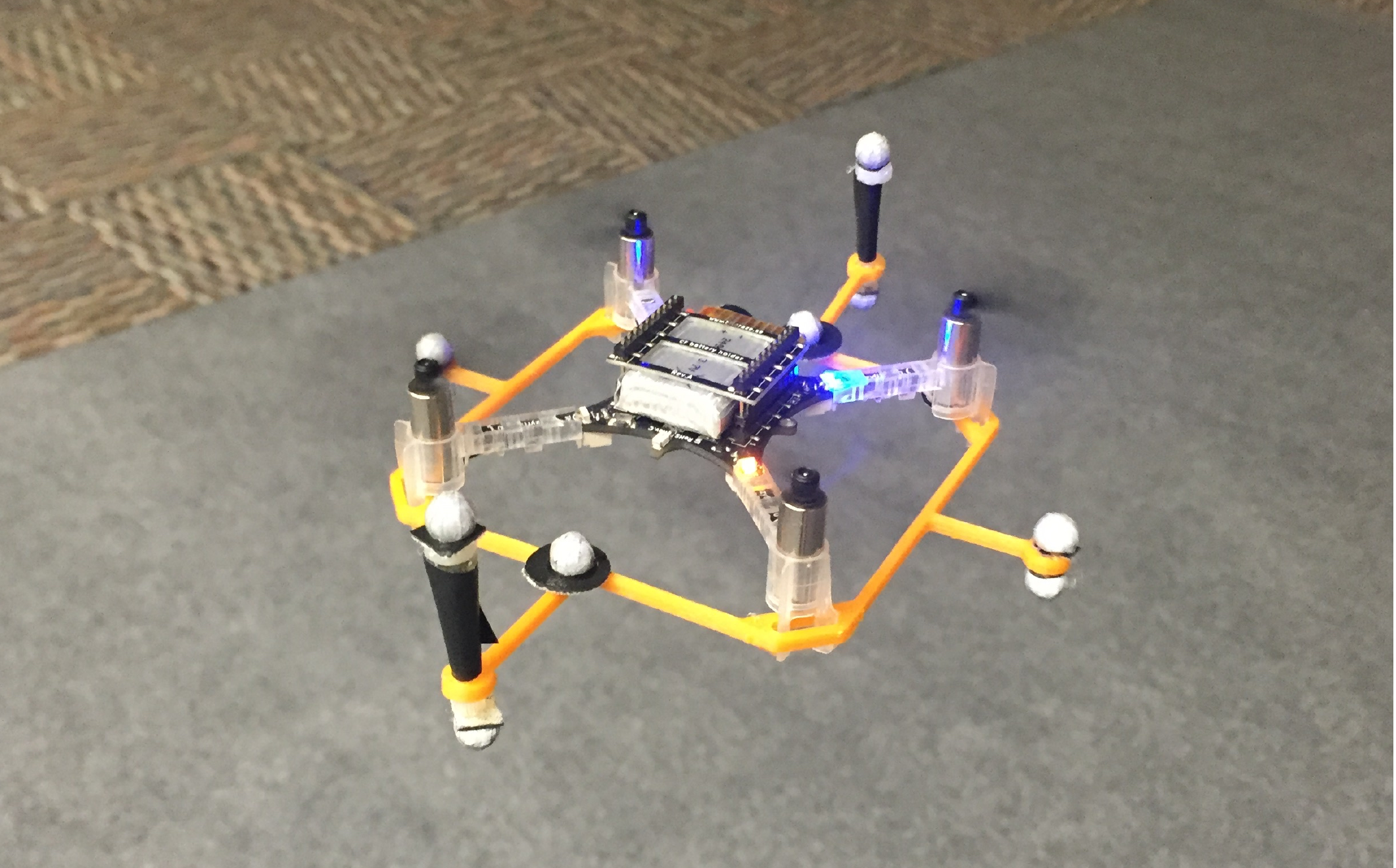}
  \caption{The Crazyflie 2.0}
  \label{fig:crazyflie}
  \vspace{-10pt}
\end{wrapfigure}
\begin{figure}[t]
  \centering
  \includegraphics[width=0.65\columnwidth]{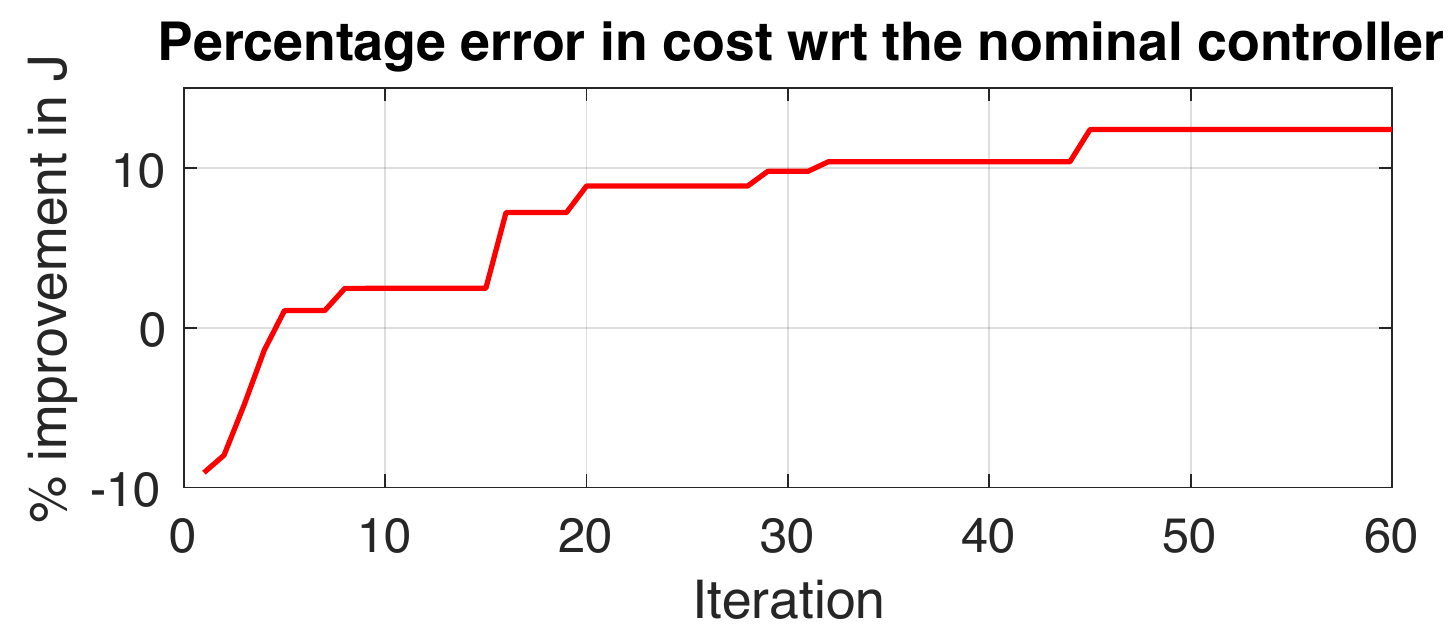}
  \caption{Crazyflie: percentage error between the learned and the nominal controller. The nominal controller is obtained by using the full 12D non-linear dynamics model of the quadrotor. As learning progresses, \metName{} outperforms the nominal controller by 12\% on the actual system, indicating the capability of \metName{} to overcome modeling errors.}
  \label{fig:exp_cost}
  \vspace{-6pt}
\end{figure}

% \caption{The Cart-pole and the Crazyflie systems}
% \label{fig:systems}
% \end{figure}
%
We now present the results of our experiments on Crazyflie 2.0, which is an open source nano quadrotor platform developed by Bitcraze. Its small size, low cost, and robustness make it an ideal platform for testing new control paradigms. Recently, it has been extensively used to demonstrate aggressive flights \cite{landry2015planning, bansal2016learning}. 
For small yaw, the quadrotor system is modeled as a rigid body with a ten dimensional state vector $s\coloneqq \begin{bmatrix} p, v, \zeta, \omega \end{bmatrix}$, which includes the position $p=(x,y,z)$ in an inertial frame $I$, linear velocities $v=(v_x,v_y,v_z)$ in $I$, attitude (orientation) represented by Euler angles $\zeta$, and angular velocities $\omega$. The system is controlled via three inputs $u \coloneqq \begin{bmatrix}u_1, u_2, u_3 \end{bmatrix}$, where $u_1$ is the thrust along the $z$-axis, and $u_2$ and $u_3$ are rolling, pitching moments respectively. The full non-linear dynamics of a quadrotor are derived in \cite{abas2011}, and its physical parameters are computed in \cite{landry2015planning}.
%\begin{equation} \label{eqn:quad_dynamics}
%\dot{s} 
%= 
%\begin{bmatrix}
%\dot{x} \\
%\dot{y} \\
%\dot{z} \\
%\dot{v_x} \\
%\dot{v_y} \\
%\dot{v_z} \\
%\dot{\phi} \\
%\dot{\psi} \\
%\dot{\omega_x} \\
%\dot{\omega_y}
%\end{bmatrix}
%=
%\begin{bmatrix}
%v_x\\
%v_y\\
%v_z\\
%-\cos\phi \sin\psi \tfrac{u_1}{m} \\
%\sin\phi \tfrac{u_1}{m} \\
%g - \cos\phi \cos\psi \tfrac{u_1}{m}  \\
%\omega_x + \sin\phi \tan\psi \omega_y\\
%\cos\phi \omega_y\\
%\tfrac{L}{I_x} u_2 \\
%\tfrac{L}{I_y} u_3
%\end{bmatrix}
%\end{equation}
%%
%where $L, m, I_x, I_y$ are physical parameters of the quadrotor and are obtained from \cite{landry2015planning}. 
Our goal in this experiment is to track a desired position $p^*$ starting from the initial position~$p_0 = [0, 0, 1]$. 
Formally, we minimize
\begin{equation} \label{eqn:costExp}
J_0(\bold{\bar{s}}_0^N, \ctrlseq_0^{N-1})= \sum_{k=0}^{N-1}\left( \bar{s}_k^T Q \bar{s} + \ctrl_k^T R \ctrl_k \right) + \bar{s}_N^T Q_f \bar{s}\,,
\end{equation}
where $\bar{s}\coloneqq \begin{bmatrix} p-p^*, v, \zeta, \omega \end{bmatrix}$. Given the dynamics in \cite{abas2011}, the desired optimal control problem can be solved using LQR; however, the resultant controller may still not be optimal for the actual system because (a) true underlying system is non-linear and (b) the actual system may not follow the dynamics in \cite{abas2011} exactly due to several unmodeled effects, as illustrated in our results. Hence, we assume that the dynamics of $v_x$ and $v_y$ are unknown, and model them as
\begin{equation} \label{eqn:linquad_dynamics}
\begin{bmatrix} f_{v_x} \\ f_{v_y} \end{bmatrix} = \paramA{} \begin{bmatrix} \phi \\ \psi \end{bmatrix} +  \paramB{} u_1\,,
\end{equation}
where $A$ and $B$ are parameterized through $\param{}$. Our goal is to learn the parameter $\param^*$ that minimizes the cost in \eqref{eqn:costExp} for the \textit{actual} Crazyflie using \metName{}. 
We use $N = 400$; the penalty matrix $Q$ is chosen to penalize the position deviation. In our experiments, Crazyflie was flown in presence of a VICON motion capture system, which along with on-board sensors provides the full state information at 100Hz. The optimal control problem for a particular linearization in \eqref{eqn:linquad_dynamics} is solved using LQR. For comparison, we compute the \textit{nominal} optimal controller using the full dynamics in \cite{abas2011}. 
Figure~\ref{fig:exp_cost} shows the performance of the controller from \metName{} compared with the nominal controller during the learning process. The nominal controller outperforms the learned controller initially, but within a few iterations, \metName{} performs better than the controller derived from the known dynamics model of Crazyflie. This is because \metName{} optimizes controller based on the performance of the \textit{actual} system and hence can account for unmodeled effects. In 45 iterations, the learned controller outperforms the nominal controller by 12\%, demonstrating the performance potential of \metName{} on real systems.
% Figure \ref{fig:exp_trajs} shows the error trajectories (from $p^*$) of $x$, $y$ and $z$ for the learned controller and the nominal controller. Although the $x$ and $y$ tracking is similar for the both controllers, the actual system is able to track the desired $z$ position much better with the learned controller, resulting in its lower cost compared to the nominal controller.   
% \begin{figure}[t]
%   \centering
%   \includegraphics[width=0.4\textwidth]{"figs/exp_trajs"}
%   \caption{Crazyflie: tracking error for the learned and nominal controllers. The final learned controller does a better tracking of $z$ state, which results in an overall better performance.}
%   \label{fig:exp_trajs}
% \end{figure}
% Experiments
%
% !TEX root = SysIdLQR.tex
\section{Conclusions and Future Work}
In this work, we introduce \metName{}, an active learning framework to optimize the system dynamics with the intent of maximizing the controller performance. Through simulations and real-world experiments, we demonstrate that \metName{} achieves the optimal control performance even when no prior information is known about the system dynamics. 

Several interesting future directions emerge from this work. 
% Generalization to the class of tasks
The dynamics model learned through \metName{} maximizes the performance on a single task. 
The obtained dynamics model thus may not necessarily perform well on a similar but different task.
It will be interesting to generalize \metName{} to optimize the dynamics for a class of tasks, e.g., regulating to different states.
% Incorporate state and trajectory data 
Leveraging the state and input trajectory data, along with the observed performance, to further increase the data-efficiency of the learning process is another promising direction.
% Safety of BO
During the learning process, \metName{} can query parameters which might lead to an unstable behavior on the actual system and can cause safety concerns. 
In such cases, it might be desirable to combine \metName{} with the exploration methods that explicitly take safety into account, e.g., such as SafeOpt~\cite{sui2015safe, berkenkamp2016safe}.  
% Scalability of BO
Finally, since BO is not scalable to higher-dimensional systems (roughly beyond 30-40 parameters) \cite{Shahriari2016}, exploring alternative ways to scale \metName{} to more complex non-linear dynamics models is another interesting direction.

\bibliographystyle{IEEEtran}
\bibliography{references}
\end{document}